\documentclass[10pt, conference]{IEEEtran}

\IEEEoverridecommandlockouts
\usepackage{cite}
\usepackage{amsmath,amssymb,amsfonts}
\usepackage{algorithmic}
\usepackage{graphicx}
\usepackage{textcomp}
\usepackage{xcolor}
\usepackage{pgfplots, filecontents}
\pgfplotsset{compat=1.18}
\usepgfplotslibrary{patchplots}
\usepackage{enumerate}
\usepackage{multirow}
\usepackage[nohyperlinks, nolist]{acronym}
\usepackage{epstopdf}
\usepackage{float}
\usepackage[caption = false]{subfig}
\usetikzlibrary{matrix}
\graphicspath{{./fig/}}
\usepackage{hyperref}
\hypersetup{hidelinks}
\usepackage{algorithm,algorithmic}

\usepackage{microtype}
\usepackage{svg}
\usepackage{amsmath,blkarray,booktabs, bigstrut}
\DeclareMathSizes{10}{8.9}{6.5}{4.5}
\everymath{\medmuskip=1.5mu minus 1.5mu\thickmuskip=2mu minus 2mu}

\usepackage{tikz, pgfplots}
\usetikzlibrary{positioning}
\usetikzlibrary{arrows}
\usetikzlibrary{shapes.multipart}
\usetikzlibrary{calc,through,backgrounds}
\newcommand{\ve}{\mathbf}
\newcommand{\m}{\mathbf}
\newcommand*{\field}[1]{\mathbb{#1}}%  
  
\definecolor{ao}{rgb}{0.0, 0.5, 0.0}
  
\newcommand*{\ReadOutElement}[4]{%
    \pgfplotstablegetelem{#2}{#3}\of{#1}%
    \let#4\pgfplotsretval
}

\def\BibTeX{{\rm B\kern-.05em{\sc i\kern-.025em b}\kern-.08em
    T\kern-.1667em\lower.7ex\hbox{E}\kern-.125emX}}
\begin{document}
\begin{acronym}

\acro{aaf}[AAF]{anti-aliasing filter}
\acro{ad}[AD]{autonomous driving}
\acro{adc}[ADC]{analog-to-digital converter}
\acro{adas}[ADAS]{advanced driver assistance systems}
\acro{als}[ALS]{approximate least squares}
\acro{anls}[ANLS]{approximate nonlinear least squares}
\acro{aoa}[AOA]{angle of arrival}
\acro{aod}[AOD]{angle of departure}
\acro{awgn}[AWGN]{additive white Gaussian noise}
\acro{bb}[BB]{baseband}
\acro{ber}[BER]{bit error ratio}
\acro{blue}[BLUE]{best linear unbiased estimator}
\acro{bmse}[BMSE]{Bayesian mean square error}
\acro{bpsk}[BPSK]{binary phase shift keying}
\acro{bwlue}[BWLUE]{best widely linear unbiased estimator}
\acro{ccdf}[CCDF]{complementary cumulative distribution function}
\acro{cfo}[CFO]{carrier frequency offset}
\acro{cfr}[CFR]{channel frequency response}
\acro{cir}[CIR]{channel impulse response}
\acro{cpe}[CPE]{common phase error}
\acro{cr}[CR]{corner reflector}
\acro{crlb}[CRLB]{Cramér-Rao lower bound}
\acro{cs}[CS]{compressed sensing}
\acro{cwcu}[CWCU]{component-wise conditionally unbiased}
\acro{cl}[CWCU LMMSE]{component-wise conditionally unbiased linear minimum mean square error}
\acro{cp}[CP]{cyclic prefix}
\acro{cwl}[CWCU WLMMSE]{component-wise conditionally unbiased widely linear minimum mean square error}
\acro{dac}[DAC]{digital-to-analog converter}
\acro{da}[DA]{Doppler aliasing}
\acro{dbf}[DBF]{digital beamforming}
\acro{dc}[DC]{direct current}
\acro{dft}[DFT]{discrete Fourier transform}
\acro{dtft}[DTFT]{discrete-time Fourier transform}
\acro{ddm}[DDM]{Doppler-division multiplexing}
\acro{dpsk}[DPSK]{differential phase shift keying}
\acro{ecir}[ECIR]{effective channel impulse response}
\acro{ecfr}[ECFR]{effective channel frequency response}
\acro{em}[EM]{expectation-maximization}
\acro{esi}[ESI]{equidistant subcarrier interleaving}
\acro{fim}[FIM]{Fisher information matrix}
\acro{fmcw}[FMCW]{frequency-modulated continuous wave}
\acro{fft}[FFT]{fast Fourier transform}
\acro{fir}[FIR]{finite impulse response}
\acro{gb}[GB]{guard band}
\acro{hpf}[HPF]{high-pass filter}
\acro{ici}[ICI]{inter-carrier interference}
\acro{iir}[IIR]{infinite impulse response}
\acro{isi}[ISI]{inter-symbol interference}
\acro{idft}[IDFT]{inverse discrete Fourier transform}
\acro{ifft}[IFFT]{inverse fast Fourier transform}
\acro{iid}[i.i.d.]{independent and identically distributed}
\acro{llr}[LLR]{log-likelihood ratio}
\acro{lmmse}[LMMSE]{linear minimum mean square error}
\acro{lms}[LMS]{least mean square}
%\acro{ln}[LN]{least norm}
\acro{lna}[LNA]{low-noise amplifier}
\acro{los}[LOS]{line of sight}
\acro{lpi}[LPI]{low probability of intercept}
\acro{ls}[LS]{least squares}
\acro{lti}[LTI]{linear time-invariant}
\acro{map}[MAP]{maximum a posteriori}
\acro{mimo}[MIMO]{multiple-input multiple-output}
\acro{miso}[MISO]{multiple-input single-output}
\acro{ml}[ML]{maximum likelihood}
\acro{mhus}[MHUS]{multiharmonic unknown signal}
\acro{mlem}[ML-EM]{maximum likelihood expectation-maximization}
\acro{mmse}[MMSE]{minimum mean square error}
\acro{mse}[MSE]{mean square error}
\acro{mvdr}[MVDR]{minimum variance distortionless response}
\acro{mvu}[MVU]{minimum variance unbiased}
\acro{nlms}[NLMS]{normalized least mean squares}
\acro{nlos}[NLOS]{non-line of sight}
\acro{ofdm}[OFDM]{orthogonal frequency-division multiplexing}
\acro{otfs}[OTFS]{orthogonal time frequency space}
\acro{papr}[PAPR]{peak-to-average power ratio}
\acro{pdf}[PDF]{probability density function}
\acro{pmf}[PMF]{probability mass function}
\acro{pll}[PLL]{phase-locked loop}
\acro{ppus}[PPUS]{periodic pulses unknown signal}
\acro{ppks}[PPKS]{periodic pulses known signal}
\acro{pri}[PRI]{pulse repetition interval}
\acro{pslr}[PSLR]{peak-to-sidelobe ratio}
\acro{psnr}[PSNR]{pseudo signal-to-noise ratio}
%\acro{pmf}[PMF]{probability mass function}
\acro{pwcu}[PWCU]{part-wise conditionally unbiased}
\acro{pwl}[PWCU WLMMSE]{part-wise conditionally unbiased widely linear minimum mean square error}
\acro{qam}[QAM]{quadrature amplitude modulation}
\acro{qpsk}[QPSK]{quadrature phase-shift keying}
\acro{rcs}[RCS]{radar cross section}
\acro{rf}[RF]{radio frequency}
\acro{rls}[RLS]{recursive least squares}
\acro{rvm}[RDM]{range-Doppler map}
%\acro{rdm}[RDM]{range-Doppler map}
%\acro{rdm}[RDMult]{range-division multiplexing}
\acro{sa}[SA]{subcarrier aliasing}
\acro{sfdr}[SFDR]{spurious-free dynamic range}
\acro{sim}[SIM]{spectral interleaving multiplexing}
\acro{siso}[SISO]{single-input single-output}
\acro{snr}[SNR]{signal-to-noise ratio}
\acro{salsa}[SALSA]{split augmented Lagrangian shrinkage}
\acro{stln}[STLN]{structured total least norm}
\acro{stls}[STLS]{structured total least squares}
\acro{tls}[TLS]{total least squares}
\acro{tdm}[TDM]{time-division multiplexing}
\acro{ula}[ULA]{uniform linear array}
\acro{uw}[UW]{unique word}
\acro{uwofdm}[UW-OFDM]{unique word orthogonal frequency division multiplexing}
%\acro{wgn}[WGN]{white Gaussian noise}
\acro{wlan}[WLAN]{wireless local area network}
\acro{wlls}[WLLS]{widely linear least squares}
\acro{wlmmse}[WLMMSE]{widely linear minimum mean square error}
\acro{wls}[WLS]{weighted least squares}
\acro{wwlls}[WWLLS]{weighted widely linear least squares}

\end{acronym}

\title{Estimators and Performance Bounds for Short Periodic Pulses \\
% \thanks{

%     }
}

\author{\IEEEauthorblockN{Sebastian~Schertler\IEEEauthorrefmark{1}\IEEEauthorrefmark{4}, Oliver~Lang\IEEEauthorrefmark{1}\IEEEauthorrefmark{4}, Jonas~Lindenberger\IEEEauthorrefmark{2}, Stefan~Schuster\IEEEauthorrefmark{3}, \\
Stefan~Scheiblhofer\IEEEauthorrefmark{3}, Alexander~Haberl\IEEEauthorrefmark{3}, Clemens~Staudinger\IEEEauthorrefmark{3}, and Mario~Huemer\IEEEauthorrefmark{1}}
\IEEEauthorblockA{\IEEEauthorrefmark{1}Institute of Signal Processing, Johannes Kepler University, 4040 Linz, Austria\\
\IEEEauthorrefmark{2}Silicon Austria Labs GmbH, 4040 Linz, Austria\\
\IEEEauthorrefmark{3}voestalpine Stahl GmbH, 4020 Linz, Austria\\
\IEEEauthorrefmark{4}The authors contributed equally to this work.
}}

\maketitle

\begin{abstract}
In many industrial applications, signals with short periodic pulses, caused by repeated steps in the manufacturing process, are present, and their fundamental frequency or period may be of interest. Fundamental frequency estimation is in many cases performed by describing the periodic signal as a multiharmonic signal and employing the corresponding maximum likelihood estimator.  
However, since signals with short periodic pulses contain a large number of noise-only samples, the multiharmonic signal model is not optimal to describe them.

In this work, two models of short periodic pulses with known and unknown pulse shape are considered. For both models, the corresponding maximum likelihood estimators, Fisher information matrices, and approximate Cram{\'e}r-Rao lower bounds are presented. 
Numerical results demonstrate that the proposed estimators outperform the maximum likelihood estimator based on the multiharmonic signal model for low signal-to-noise
ratios.
\end{abstract}

\begin{IEEEkeywords}
CRLB, periodic signal, maximum likelihood, fundamental frequency, non-linear least squares, pitch estimation
\end{IEEEkeywords}

\section{Introduction}  \label{sec:Introduction}

\IEEEPARstart{P}{eriodic} signals appear in many practical applications and are frequently modeled as a multiharmonic signal with spectral components at the fundamental frequency and its integer multiples \cite{hess2012pitch, 6850156, Wise1976}. If the task is to estimate the signal's fundamental frequency or it's corresponding period length, frequently the \ac{ml} estimator is applied \cite[p.~292ff.]{kay2017fundamentals}. 

\begin{figure}[!t]
\centering
\begin{tikzpicture}[scale=0.8]
% Draw axes
    \draw [<->,thick] (0,1.8) node (yaxis) [above] {}
        |- (8.5,0) node (xaxis) [below] {};    
    \node[below] at (0,0 ) {\footnotesize $0$}; 
    \node[above] at (8.5,0 ) {\small $t$}; 
    \node[left, anchor=south, rotate=90] at (0,0.7) {\small Signal};
    %\node[below] at (1.5,0) {\footnotesize 1023};
    %\node[below] at (3,0) {\footnotesize 2047};
    \node[below] at (7.5,0) {\footnotesize $(N-1)T_\text{s}$};
    \draw [black] plot coordinates { (7.5,0.2) (7.5,-0.2) };
    
    %\node[below] at (3,-0.5) {Time domain samples};

% Draw colors
%	\draw[draw=none, fill=gray, opacity=0.2] (0,0) rectangle ++(1.5, 2);
%	\draw[draw=none, fill=gray, opacity=0.4] (1.5,0) rectangle ++(1.5, 2);
%	\draw[draw=none, fill=gray, opacity=0.2] (3,0) rectangle ++(1.5, 2);
%	\draw[draw=none, fill=gray, opacity=0.4] (4.5,0) rectangle ++(1.5, 2);
	
% CIR
\draw [black, xshift=1.0cm] plot [smooth, tension=1] coordinates { (0,0) (0.15,1.0) (0.3,0) };

\draw [black, xshift=2.5cm] plot [smooth, tension=1] coordinates { (0,0) (0.15,1.0) (0.3,0) };

\draw [black, xshift=4.0cm] plot [smooth, tension=1] coordinates { (0,0) (0.15,1.0) (0.3,0) };

\draw [black, xshift=5.5cm] plot [smooth, tension=1] coordinates { (0,0) (0.15,1.0) (0.3,0) };

\draw [black, xshift=7.0cm] plot [smooth, tension=1] coordinates { (0,0) (0.15,1.0) (0.3,0) };

% Arrows
\draw [<->,thick] (0,1.2)--(1,1.2) node[midway,above, xshift=0.3cm] () {\footnotesize $\tau_0 \approx n_0 T_\text{s}$}; 

\draw [<->,thick] (2.5,1.2)--(4,1.2) node[midway,above, xshift=0.0cm] () {\footnotesize $T \approx P T_\text{s}$};

\draw [<->,thick] (5.5,1.2)--(5.8,1.2) node[midway,above, xshift=0.0cm] () {\footnotesize $T_\text{p} \approx N_\text{p} T_\text{s}$}; 

%\draw[decorate,decoration={brace,amplitude=5pt}, thick]  (0,2.3) -- (0.5,2.3);
%\node at (0.25, 2.8) {CIR $\ve{f}_0$};
%\node at (2, 2.8) {$\hdots$};
%\draw[decorate,decoration={brace,amplitude=5pt}, thick]  (0,3.2) -- (6,3.2);
%\node (s1) at (3, 3.7) {ECIR $\ve{g}$};

% Tx indication
%\node (h0) at (0.75, 1.75) {Tx0};
%\node (h1) at (2.25, 1.75) {Tx1};
%\node (h2) at (3.75, 1.75) {Tx2};
%\node (h3) at (5.25, 1.75) {Tx3};

\end{tikzpicture}
\caption{Exemplary signal with short periodic pulses with period length $T$, pulse duration $T_\text{p}$, delay of the first pulse $\tau_0$, and sampling interval $T_\text{s}$.}
\label{fig:pulsed_signal}
\end{figure}
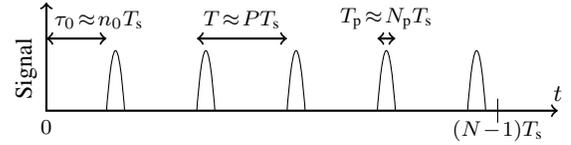

However, in many applications, the multiharmonic model is not the best choice to describe the signal, examined in this work. 
The type of signals considered in this work (exemplarily sketched in Fig.~\ref{fig:pulsed_signal})  are periodically appearing short pulses whose lengths are much shorter than the period length. 
Consequently, many signal samples are zero and the corresponding measurements contain noise only. Pulsed signals can be modeled approximately as multiharmonic. However, this would not reflect the short pulse duration. In case the pulse length is known, the question arises of whether this knowledge can be used in a beneficial way to increase the performance of the estimated period length compared to the \ac{ml} estimator based on the multiharmonic signal.

In this work, we consider two different signal models. For the first one, the pulse shape is assumed to be known up to an unknown but constant scaling factor. This model may be employed in applications where a known impulse response is triggered periodically (by a delta-peak-like trigger signal). For the second model, the pulse duration is assumed to be known, but the pulse shape is unknown and will be estimated jointly with the period length. For each model, this work derives the corresponding \ac{ml} estimator and the \ac{fim}.

To the best knowledge of the authors, these results have not been reported in the literature so far. Related literature includes \cite[p.~71ff.]{kay2017fundamentals} describing a time-domain model and algorithms for periodic signals, but short pulses (compared to the period length) were not considered. Related literature also includes \cite{bang2023accurate}, where binary modulated radar pulse trains are considered as \ac{lpi} signals in electronic warfare support, and several common pulse shapes are investigated. The authors present a method for estimating the \ac{pri} and other waveform parameters based on a change point detection algorithm, whose purpose is to detect the on- and off-set of pulses. In \cite{kumar2014deinterleaving, albaker2009signal}, the \ac{pri} interval of a radar pulse train is estimated based on envelope detectors. 
Another related literature is \cite{nishiguchi2000improved}, where the \ac{pri} interval of a radar pulse train is estimated based on a \ac{pri} transform \cite{nishiguchi1983new, deng2012sfm} using a combinatorial algorithm. The strength of this algorithm is that it suppresses the subharmonics in the autocorrelation
function almost completely. Other autocorrelation-based algorithms to estimate \ac{pri} of pulse trains include \cite{mardia1989new, milojevic1992improved}. An approximate \ac{ml} period length estimator for a signal model containing the time indices of the occurring pulses is presented in \cite{clarkson2008approximate}, which also allows for missing pulses. Some literature considers binary (1-bit quantized) pulses, for example, with the goal of identifying/deinterleaving several sources of such pulses as in \cite{logothetis1998interval}.

 In this work, \ac{ml} estimators and the \acp{fim} are derived for both considered models of short periodic pulses. It will turn out that the approximate \ac{fim} for the second model is singular and therefore cannot be inverted to obtain the \ac{crlb}, which will be discussed. In addition, it will turn out that the \ac{crlb} for estimating the period length based on the first signal model approximately coincides with that obtained with the multiharmonic model. Hence, asymptotically, the \ac{ml} estimators cannot beneficially utilize the knowledge of the short pulse duration. However, in the threshold region \cite{Williamson1994}, both \ac{ml} estimators based on the introduced signal models outperform the multiharmonic estimator. The estimator based on the first signal model also shows better results in the low to moderate \ac{snr} region.

\section{Signal Models}  \label{sec:Models}

\subsection{Multiharmonic Signal Model}
The multiharmonic signal model

\begin{align}
	x[n] ={}& \sum_{k=1}^{K_\text{h}} A_k\, \text{sin}( 2\pi k \psi_0  n + \phi_k) + w[n] \label{equ:MultHarmModel}
\end{align}

includes the normalized fundamental frequency or normalized pitch $\psi_0$, the amplitudes $A_k$ and phases $\phi_k$ of the $K_\text{h}$ spectral components. $T_\text{s}$ is the sampling interval at which $N$ measurements are gathered, and  $w[n] \sim \mathcal{N}(0, \sigma^2)$ is \ac{awgn} with $0 \leq n < N,\,\, n \in \field{N}$. If only $K_\text{h}$ is known and the normalized fundamental frequency $\psi_0$ is to be estimated, the \ac{crlb} is approximated by \cite{NIELSEN2017188,nehorai1986adaptive}
\begin{align}
	\text{var}(\widehat{\psi}_0) \geq{}&  2 \sigma^2 \frac{3}{N(N^2-1) \pi^2\, \sum_{k=1}^{K_\text{h}} k^2\,A_k^2} \label{equ:FD_CRLB_psi}.
\end{align}	
For $K_\text{h} \psi_0 < \frac{1}{2}$, the \ac{ml} estimate for $\psi_0$ can be found by a grid search \cite[p.~293]{kay2017fundamentals}

\begin{align}
    \widehat{\psi}_0 ={}&  \text{arg}\, \underset{\psi_0}{\text{max}} \,\, \ve{x}^T  \m{H}(\psi_0) \left(\m{H}(\psi_0)^T \m{H}(\psi_0)\right)^{-1} \m{H}(\psi_0)^T  \ve{x}, \label{equ:FD_Est_ML}
\end{align}

where $\ve{x}$ is a column vector of length $N$ containing the measurements $x[n]$.   This estimator is simply referred to as "\ac{mhus} \ac{ml} estimator" in the simulation results in Sec.~\ref{sec:Simulations}.  $\m{H}(\psi_0 )$ is given by

\begin{align}
	\m{H}(\psi_0 ) ={}& \begin{bmatrix}
	\text{cos}(2\pi \psi_0 0) & \hdots & \text{cos}(2\pi \psi_0 (N-1) ) \\
	\vdots & & \vdots \\	
	\text{cos}(2\pi K_\text{h} \psi_0 0) & \hdots & \text{cos}(2\pi K_\text{h} \psi_0 (N-1)) \\	
	\text{sin}(2\pi \psi_0 0) & \hdots & \text{sin}(2\pi \psi_0 (N-1)) \\
	\vdots & & \vdots \\
	\text{sin}(2\pi K_\text{h} \psi_0 0) & \hdots & \text{sin}(2\pi K_\text{h} \psi_0 (N-1) )
	\end{bmatrix}^T.
\end{align}

Under the condition $0\ll \psi_0 \ll \frac{1}{2K_\text{h}} $, a computationally efficient approximation of the \ac{ml} estimator, called the \ac{anls} fundamental frequency estimator, is shown in \cite{10476910} to be
\begin{align}
    \widehat{\psi}_0 ={}&  \text{arg}\, \underset{\psi_0}{\text{max}} \,\, \sum_{k=1}^{K_\text{h}} |X(k \psi_0)|^2, \label{equ:FD_Est_psi}
\end{align}	

with the windowed \ac{dtft} $X(\psi_0) = \sum_{n=0}^{N-1} x[n] \text{e}^{-\text{j} 2\pi \psi_0 n}$ that can be efficiently evaluated using the \ac{fft} \cite{CooleyTukey}.
This estimator is simply referred to as "\ac{mhus} \ac{anls} estimator" in the simulation results in Sec.~\ref{sec:Simulations}.
An estimate for the period length $T = \frac{T_\text{s}}{\psi_0}$ can simply be found by $\widehat{T} = \frac{T_\text{s}}{\widehat{\psi}_0}$ with the transformed \ac{crlb} \cite[p.~45]{Kay-Est.Theory} given by
\begin{align}
	\text{var}(\widehat{T}) \geq{}& \frac{6 \sigma^2 T^4}{\pi^2 N(N^2-1) T_\text{s}^2 \sum_{k=1}^{K_\text{h}} k^2\,A_k^2}. \label{equ:FD_CRLB_T}
\end{align}

\subsection{Signal Models with Short Periodic Pulses}

In this work, we consider two different signal models for short periodic pulses as indicated in Fig.~\ref{fig:pulsed_signal}. For the first model, pulses with known shape $s_\text{ps}(t)$ but unknown signal scale $A$ are repeated with a period length $T$. The pulses have a known length of $T_\text{p}$ and $s_\text{ps}(t)=0$ for $t<0$ and $t>T_\text{p}$. The $N$ measurements $x[n]$ can be modeled as $x[n] = s_1(n T_\text{s}; \boldsymbol\theta_1) + w[n]$ with
\begin{align}
	s_1(t; \boldsymbol\theta_1)={}&\sum_{k=-\infty}^{\infty} A s_\text{ps}(t - k T  - \tau_0 ), \label{equ:TDModel1}
\end{align}
where $0 \leq n < N, n \in \field{N}$, $w[n] \sim \mathcal{N}(0, \sigma^2)$,  and where $\tau_0 < T$ is the time delay of the first pulse w.r.t. the first measurement. The parameter vector is given by $\boldsymbol\theta_1 = [T, \tau_0, A]^T$. 

For the second model, the signal's shape is assumed to be unknown, allowing for dismissing the signal scale $A$ in the model, leading to $x[n] = s_2(n T_\text{s}; \boldsymbol\theta_2) + w[n]$ with
\begin{align}
	 s_2(t; \boldsymbol\theta_2) ={}& \sum_{k=-\infty}^{\infty} s_\text{p}(t - k T  - \tau_0 ), \label{equ:TDModel2}
\end{align}
where $s_\text{p}(t)$ is the unknown pulse with known length $T_\text{p}$ and with $s_\text{p}(t)=0$ for $t<0$ and $t>T_\text{p}$. Although estimating the period length $T$ is of primary focus, utilizing a separable \ac{ls} approach on the model in \eqref{equ:TDModel2} requires estimating the pulse $s_\text{p}(t)$ (as will be done in Sec.~\ref{sec:Estimators}). Using $s_\text{p}[n] = s_\text{p}(n T_\text{s})$, the parameter vector of length $N_\text{p}+2$ with $N_\text{p} = \lceil T_\text{p}/T_\text{s}\rceil$ is $\boldsymbol\theta_2 = [T, \tau_0, s_\text{p}[0], \hdots s_\text{p}[N_\text{p}-1]]^T$.

\section{Cramèr-Rao-Lower-Bound}  \label{sec:CRLB}

In this section, the \acp{fim} for both models and approximations of them that enable closed-form analytical expressions are derived. The latter can be inverted for the first signal model, yielding a closed-form expression of the \ac{crlb} \cite{MR15748, cramér1946mathematical} of the estimated period length $\widehat{T}$.

\subsection{FIM Derivation}

The \ac{fim} for the signals in \ac{awgn} \cite[p.~49]{Kay-Est.Theory} combined with the signal model in \eqref{equ:TDModel1} with known pulse shape can be rewritten as a sum of matrices (using a numerator layout)
\begin{align}
	\m{I}_1(\boldsymbol\theta_1) ={}& \frac{1}{\sigma^2} \sum_{n=0}^{N-1}  \left[ \frac{\partial s_1(n T_\text{s}; \boldsymbol\theta_1)}{\partial \boldsymbol\theta_1} \right]^T \left[ \frac{\partial s_1(n T_\text{s}; \boldsymbol\theta_1)}{\partial \boldsymbol\theta_1} \right]. \label{equ:CRLB_01a}
\end{align}
The required partial derivatives are

\begin{align}
	\frac{\partial s_1(n T_\text{s}; \boldsymbol\theta_1)}{\partial T} ={}& -A\sum_{k=-\infty}^{\infty}  \frac{\partial s_\text{ps}(t) }{\partial t} \bigg|_{t=n T_\text{s} - k T  - \tau_0 } \cdot k, \label{equ:CRLB_01} \\
	\frac{\partial s_1(n T_\text{s}; \boldsymbol\theta_1)}{\partial \tau_0} ={}& - A \sum_{k=-\infty}^{\infty} \frac{\partial s_\text{ps}(t) }{\partial t} \bigg|_{t=n T_\text{s} - k T  - \tau_0 }, \label{equ:CRLB_02} \\
	\frac{\partial s_1(n T_\text{s}; \boldsymbol\theta_1)}{\partial A} ={}&  \sum_{k=-\infty}^{\infty}  s_\text{ps}(t) \big|_{t=n T_\text{s} - k T  - \tau_0 }. \label{equ:CRLB_03}
\end{align}
A combination of \eqref{equ:CRLB_01a}--\eqref{equ:CRLB_03} yields the \ac{fim} $\m{I}_1(\boldsymbol\theta_1)$ whose inversion gives the \ac{crlb} for the model in \eqref{equ:TDModel1}.
Similarly, for the model in \eqref{equ:TDModel2} with an unknown pulse we obtain 
\begin{align}
	\m{I}_2(\boldsymbol\theta_2) ={}& \frac{1}{\sigma^2} \sum_{n=0}^{N-1}  \left[ \frac{\partial s_2(n T_\text{s}; \boldsymbol\theta_2)}{\partial \boldsymbol\theta_2} \right]^T \left[ \frac{\partial s_2(n T_\text{s}; \boldsymbol\theta_2)}{\partial \boldsymbol\theta_2} \right], \label{equ:CRLB_01b}
\end{align}
with
\begin{align}
	\frac{\partial s_2(n T_\text{s}; \boldsymbol\theta_2)}{\partial T} ={}& -\sum_{k=-\infty}^{\infty}  \frac{\partial s_\text{p}(t) }{\partial t} \bigg|_{t=n T_\text{s} - k T  - \tau_0 } \cdot k, \label{equ:CRLB_04} \\
	\frac{\partial s_2(n T_\text{s}; \boldsymbol\theta_2)}{\partial \tau_0} ={}& -  \sum_{k=-\infty}^{\infty} \frac{\partial s_\text{p}(t) }{\partial t} \bigg|_{t=n T_\text{s} - k T  - \tau_0 }, \label{equ:CRLB_05}
\end{align}
For the partial derivatives of $s_2(n T_\text{s}; \boldsymbol\theta_2)$ in \eqref{equ:TDModel2} w.r.t. the samples of the discretized pulse $s_\text{p}[a]$ the continuous pulse  $s_\text{p}(t)$ is approximated as a series of rectangular windows 
%$\text{sinc}(x) = \text{sin}(\pi x) / (\pi x)$ in form of
\begin{align}	 
	s_\text{p}(t ) 	
    %\approx{}& 
    %\sum_{a'=0}^{N_\text{p}-1} s_\text{p}[a']\, \text{sinc}\left( \left( t - a' T_\text{s}\right) / T_\text{s} \right). \\
    \approx{}&  \sum_{a'=0}^{N_\text{p}-1} s_\text{p}[a'] \text{rect}\left(\frac{ t - a' T_\text{s}}{T_\text{s}} \right)
\end{align}
with $\text{rect}(t) =1 $ for $0\leq t < 1$ and 0 otherwise. Then, the partial derivative of $s_2(n T_\text{s}; \boldsymbol\theta_2)$ in \eqref{equ:TDModel2} w.r.t. $s_\text{p}[a]$ follows as
\begin{align}	
    \frac{\partial s_2(n T_\text{s}; \boldsymbol\theta_2)}{\partial s_\text{p}[a]} \approx{}& \sum_{k=-\infty}^{\infty} \text{rect}\left(\frac{n T_\text{s}- k T  - \tau_0 - a T_\text{s}}{ T_\text{s}} \right) \label{equ:CRLB_01bcc}
\end{align}
for $0 \leq a < N_\text{p}-1$. 
A combination of \eqref{equ:CRLB_01b}--\eqref{equ:CRLB_05} and \eqref{equ:CRLB_01bcc} yields the \ac{fim} $\m{I}_2(\boldsymbol\theta_2)$ for the model in \eqref{equ:TDModel2}.

\subsection{Closed-Form Approximate \Ac{fim}}

A closed-form approximation of the \acp{fim} derived previously can be obtained assuming that $K =\lfloor \frac{NT_\text{s}}{T} \rceil$ full pulses are covered within the $N$ measurements. Furthermore it is assumed that $\frac{T}{T_\text{s}}, \frac{\tau_0}{T_\text{s}} \in \field{N}$ s.t. all summands in \eqref{equ:CRLB_01bcc} evaluate to 0 or 1. Doing so yields the closed-form expression for the \ac{fim} $\m{I}_1(\boldsymbol\theta_1)$ 
\begin{align}
	& \m{I}_1(\boldsymbol\theta_1) \approx \nonumber \\
	& \frac{1}{\sigma^2} \begin{bmatrix}
	A^2 \frac{2K^3-3K^2+K}{6} \mathcal{E}^\text{ps}_\partial & A^2 \frac{K^2-K}{2}  \mathcal{E}^\text{ps}_\partial & -A \frac{K^2-K}{2}  \mathcal{S}^\text{ps}_\partial \\
	A^2 \frac{K^2-K}{2}  \mathcal{E}^\text{ps}_\partial &   A^2 K \mathcal{E}^\text{ps}_\partial & - A K  \mathcal{S}^\text{ps}_\partial \\
 - A \frac{K^2-K}{2}  \mathcal{S}^\text{ps}_\partial	& - A K  \mathcal{S}^\text{ps}_\partial &  K \mathcal{E}^\text{ps} \end{bmatrix},
 \label{equ:approx_fim1}
\end{align}	
where
\begin{align}
 \mathcal{E}^\text{ps} ={}& \sum_{n=0}^{N_\text{p}-1}  s_\text{ps}(n T_\text{s})^2 \label{equ:CRLB_07} \\
 \mathcal{E}^\text{ps}_\partial ={}& \sum_{n=0}^{N_\text{p}-1} \left(  \frac{\partial s_\text{ps}(t) }{\partial t} \bigg|_{t=n T_\text{s}} \right)^2  \label{equ:CRLB_08} \\
 \mathcal{S}^\text{ps}_\partial ={}&  \sum_{n=0}^{N_\text{p}-1} \frac{\partial s_\text{ps}(t) }{\partial t} \bigg|_{t=n T_\text{s}} s_\text{ps}(n T_\text{s}).\label{equ:CRLB_09} 
\end{align}
$\mathcal{E}^\text{ps}$ and $ \mathcal{E}^\text{ps}_\partial $ can be approximated by integrals assuming that $T_\text{s}$ is small enough \cite[p.~55]{Kay-Est.Theory}. This allows us to find a connection to the mean squared bandwidth $\overline{F^2}$ in the form
\begin{align}
	\overline{F^2} =& \frac{\frac{1}{T_\text{s}}\int_{0}^{T}\left(\frac{\partial s_\text{p}(t)}{\partial t}\right)^2\mathrm{d}t }{\frac{1}{T_\text{s}} \int_{0}^{T}  s_\text{p}(t)^2} = \frac{\mathcal{E}^\text{ps}_\partial}{\mathcal{E}^\text{ps}}, \label{equ:CRLB_12} 
\end{align}
illustrating the influence of the mean squared bandwidth $\overline{F^2}$ on the approximate \ac{fim} in \eqref{equ:approx_fim1}.
Inverting \eqref{equ:approx_fim1} leads to the \ac{crlb} for $\widehat{T}$ in form of
\begin{align}
	\text{var}(\widehat{T}) \geq{}&   \frac{12 \sigma^2 }{ (K^3 - K) A^2\mathcal{E}^\text{ps}_\partial  } \approx \frac{12 T^3 \sigma^2 }{ N^3 T_\text{s}^3 A^2 \mathcal{E}^\text{ps} \overline{F^2}  }. \label{equ:CRLB_MSB}
\end{align}

For the second signal model in \eqref{equ:TDModel2}, the closed-form approximation for the \ac{fim} $\m{I}_2(\boldsymbol\theta_2)$ is provided in \eqref{equ:CRLB_M2A}, where $\mathcal{E}^\text{p}$, $\mathcal{E}^\text{p}_\partial$, and $\mathcal{S}^\text{p}_\partial$ are defined by \eqref{equ:CRLB_07}--\eqref{equ:CRLB_09} when replacing $s_\text{ps}(t)$ by $s_\text{p}(t)$. The rather similar derivations are not shown in this work due to page limitations. 

Interestingly, the \ac{fim} in \eqref{equ:CRLB_M2A} is singular (the proof is omitted due to page limitations), due to the fact that a slightly different delay $\tau_0$ can also be represented by a slightly different estimate for the pulse $s_\text{p}[a]$ (similar to \cite{6850156}), representing an identifiability problem.
Since estimating the period length $T$ is the primary focus, it turned out that using Tikhonov regularization \cite{Hoerl01021970} prior to inverting $\m{I}_2(\boldsymbol\theta_2)$ yields approximately the same bound for estimating $T$ as obtained with the \ac{fim} $\m{I}_1(\boldsymbol\theta_1)$ for the model in \eqref{equ:TDModel1} with known pulse shape. Defining the parameter $\tau_0$ as known would also overcome the identifiability problem and yield an invertible \ac{fim}, but this scenario is unrealistic for most real-world applications. 
%whose \ac{crlb} is lower than that for the first signal model, but this scenario is unrealistic for most real-world applications.

\begin{table*}
%\caption{Derivation of Result XYZ}
\centering
\begin{minipage}{0.99\textwidth}
\begin{align}
	\m{I}_2(\boldsymbol\theta_2) \approx{}& \frac{1}{\sigma^2} \begin{bmatrix}
	 \frac{2K^3-3K^2+K}{6} \mathcal{E}^\text{p}_\partial &  \frac{K^2-K}{2}  \mathcal{E}^\text{p}_\partial & -  \frac{K^2-K}{2} \frac{\partial  s_\text{p}(0 T_\text{s})}{\partial t} & \hdots & \hdots & -  \frac{K^2-K}{2} \frac{\partial  s_\text{p}((N_\text{p}-1) T_\text{s})}{\partial t} \\	
	 \frac{K^2-K}{2}  \mathcal{E}^\text{p}_\partial &   K \mathcal{E}^\text{p}_\partial & - K \frac{\partial  s_\text{p}(0 T_\text{s})}{\partial t} & \hdots & \hdots & - K \frac{\partial  s_\text{p}((N_\text{p}-1) T_\text{s})}{\partial t} \\
-  \frac{K^2-K}{2} \frac{\partial  s_\text{p}(0 T_\text{s})}{\partial t} & -  K \frac{\partial  s_\text{p}(0 T_\text{s})}{\partial t} & K  & 0 & \hdots & 0 \\
\vdots & \vdots & 0 & K  &  & \vdots \\
\vdots & \vdots & \vdots &  &  &  0 \\
-  \frac{K^2-K}{2} \frac{\partial  s_\text{p}((N_\text{p}-1) T_\text{s})}{\partial t} & -  K \frac{\partial  s_\text{p}((N_\text{p}-1) T_\text{s})}{\partial t} & 0 & \hdots  & 0 & K
	\end{bmatrix} \label{equ:CRLB_M2A} 
\end{align}	
\medskip
\hrule
\end{minipage}
\end{table*}

\section{Maximum Likelihood Estimators} \label{sec:Estimators}

The \ac{ml} estimators for both signal models are derived in the following. Since for the considered case of \ac{awgn} with a constant noise variance for every sample, the position of the maximum of the likelihood function coincides with the position of the minimum of the \ac{ls} cost function \cite[p.~254ff.]{Kay-Est.Theory}. Thus, a separable \ac{ls} approach is used to find the \ac{ml} estimators. This can be enabled by introducing the rounded period length in samples $P = \lfloor T/T_\text{s}\rceil$ and the delay $n_0 =\lfloor \tau_0/T_\text{s}\rceil$. The estimates for $P$ derived in the following can then simply be transformed into estimates of $T \approx P T_\text{s}$. 

\subsection{ML Estimator for First Signal Model}

By assembling all $N$ measurements and noise samples into the vectors $\ve{x} = [x[0], \hdots, x[N-1]]^T$,  storing the known pulse shape in a vector $\ve{s}_\text{ps} = [s_\text{ps}[0], \hdots, s_\text{ps}[N_\text{p}-1]]^T$, separating parameters in which the model is linear and nonlinear as $\boldsymbol\alpha  = A$ and $\boldsymbol\beta = [P, n_0]^T$, respectively, one can define a model as $\ve{x} ={} \ve{h}(\boldsymbol\beta)\boldsymbol\alpha + \ve{w}$
\begin{align}
	\ve{h}(\boldsymbol\beta) ={}& [ \ve{0}_{n_0}^T,\,\, \ve{s}_\text{ps}^T,\,\, \ve{0}_{P-N_\text{p}}^T,\,\, \ve{s}_\text{ps}^T,\,\,  \ve{0}_{P-N_\text{p}}^T,\,\, \hdots ]^T,
\end{align}
where $\ve{0}_{m}$ denotes a column zero vector of length $m$. 
Inserting the \ac{ls} estimate for $\widehat{\boldsymbol\alpha}  = \left(\ve{h}(\boldsymbol\beta)^T \ve{h}(\boldsymbol\beta)\right)^{-1} \ve{h}(\boldsymbol\beta)^T  \ve{x}$, which depends on $\boldsymbol\beta$,
 into the \ac{ls} cost function results in the \ac{ml} estimate for the remaining parameters
\begin{align}
	\widehat{\boldsymbol\beta}  ={}& \text{arg}\, \underset{\boldsymbol\beta}{\text{max}} \,\, \ve{x}^T \ve{h}(\boldsymbol\beta)(\ve{h}(\boldsymbol\beta)^T \ve{h}(\boldsymbol\beta))^{-1} \ve{h}(\boldsymbol\beta)^T\ve{x}.
	\label{equ:TDKS_estim}
\end{align}
 This estimator is simply referred to as "\ac{ppks}" in the simulation results in Sec.~\ref{sec:Simulations}.
 
Several methods are possible to find the maximum of the \ac{ml} estimator's cost function, e.g., the Gauss-Newton algorithm. In this work a simple grid search is performed on the allowable range of $P$, which depends on the application, and on $0 \leq n_0 < P$.

With the assumption that approximately $K =\lfloor \frac{N}{P} \rceil$ full pulses are present in the signal, the cost function in (\ref{equ:TDKS_estim}) simplifies to
$J(\boldsymbol{\beta}) = \frac{||\ve{h}(\boldsymbol\beta)^T\ve{x}||^2}{K\mathcal{E}^\text{ps}}$.

\subsection{ML Estimator for Second Signal Model}

\begin{figure}[t]
	\centering
	\includegraphics[width=\columnwidth]{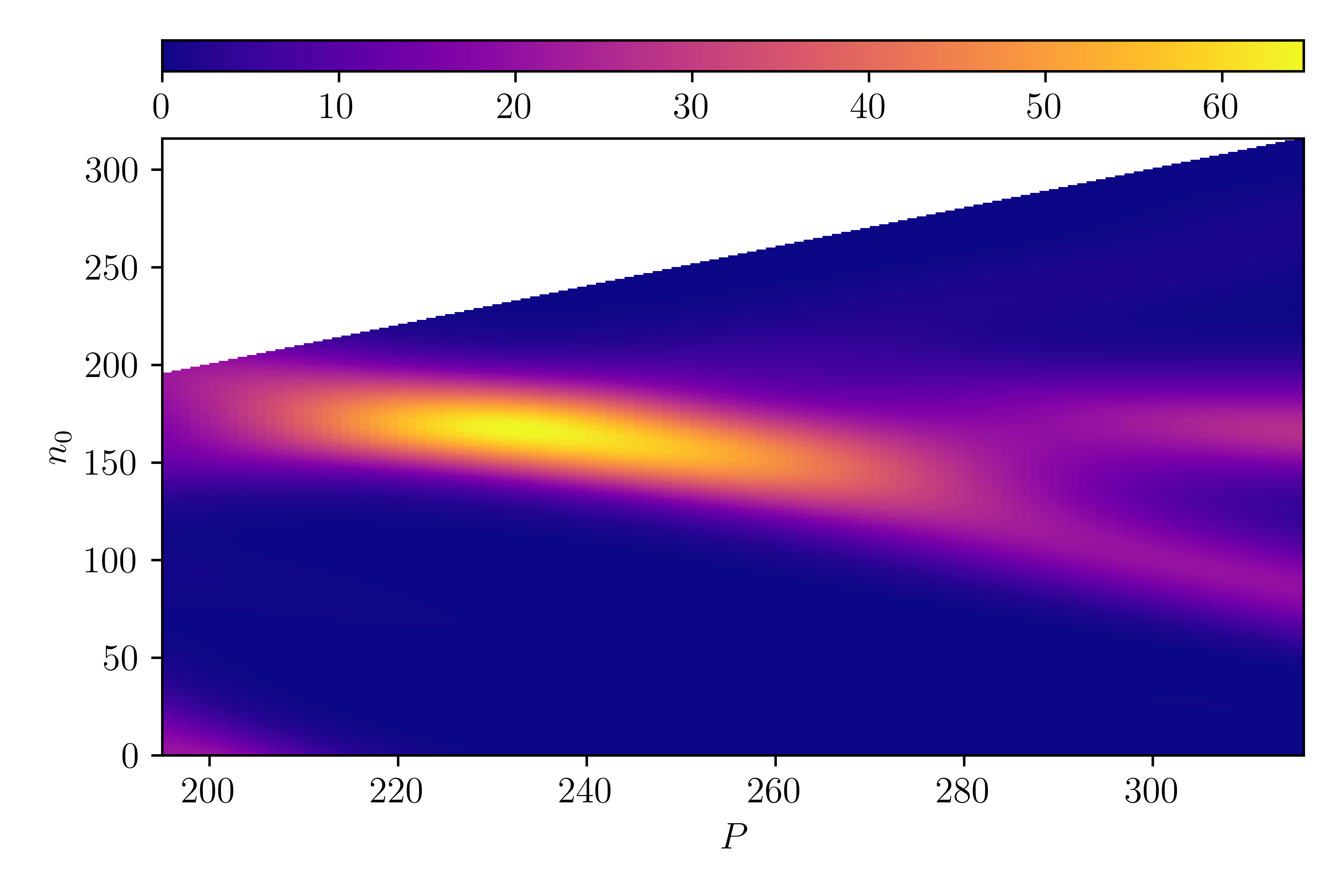}
	\caption{
		An exemplarily cost function of the ML estimator in (\ref{equ:TDKS_estim}) that depends on $P$ and $n_0$. As $n_0 < P$ invalid values have not been evaluated. 
	}
	\label{fig:costfunc}
\end{figure}

For the model in \eqref{equ:TDModel2} with unknown pulse shape one can use a similar approach. We again seperate the parameters into  $\boldsymbol\alpha  = \ve{s}_\text{p} = [s_\text{p}[0], \hdots, s_\text{p}[N_\text{p}-1]]^T$ which comprises of the unknown pulse samples and $\boldsymbol\beta = [P, n_0]^T$.
The measurement matrix $\m{H}(\boldsymbol\beta)$ is of size $N \times N_\text{p}$ is
\begin{align*}
	\m{H}(\boldsymbol\beta) ={}& [ \ve{0}_{n_0,N_\text{p}}^T,\,\, \m{I}_{N_\text{p}},\,\, \ve{0}_{P-N_\text{p},N_\text{p}}^T,\,\, \m{I}_{N_\text{p}},\,\,  \ve{0}_{P-N_\text{p},N_\text{p}}^T,\,\, \hdots ]^T,
\end{align*}
where $\ve{0}_{m,l}$ denotes a zero matrix of size $m \times l$, and $\m{I}_m$ is the identity matrix of size $m \times m$. With this measurement matrix, the measurement vector can  be described by $\ve{x} ={} \m{H}(\boldsymbol\beta)\boldsymbol\alpha + \ve{w}$.
Reinserting the \ac{ls} estimate for $\widehat{\boldsymbol\alpha}$ 
into the \ac{ls} cost function produces
\begin{align}
	\widehat{\boldsymbol\beta}  &= \text{arg}\, \underset{\boldsymbol\beta}{\text{max}} \,\, \ve{x}^T \m{H}(\boldsymbol\beta)(\m{H}(\boldsymbol\beta)^T \m{H}(\boldsymbol\beta))^{-1} \m{H}(\boldsymbol\beta)^T\ve{x},
	\label{equ:TDUS_estim}
\end{align}
which represents the \ac{ml} estimate for the second signal model. An exemplary cost function is shown in Fig. \ref{fig:costfunc}. This estimator is simply referred to as "\ac{ppus}" in the simulation results in Sec.~\ref{sec:Simulations}.
With the assumption that approximately $K =\lfloor \frac{N}{P} \rceil$ full pulses are present in the signal, the cost function in (\ref{equ:TDUS_estim}) simplifies to
$J(\boldsymbol{\beta}) = \frac{||\m{H}(\boldsymbol\beta)^T\ve{x}||^2}{K}$.

\subsection{Sub-Grid Resolution}
As the grid resolution for $\widehat{T} = \widehat{P} T_\text{s}$ is $T_\text{s}$, the \ac{mse} cannot be smaller $\frac{T_\text{s}^2}{12}$, which is the variance for a uniform distribution of width $T_\text{s}$. To receive results with a lower \ac{mse} we resample the measurement vector $\ve{x}$ by a factor of $P_\text{R}$ to virtually increase the resolution. Then, the smallest possible \ac{mse} with resampling is $\frac{(T_\text{s}/P_\text{R})^2}{12} $.

\section{Simulation Results} \label{sec:Simulations}

The sampling interval is chosen as $T_\text{s}=1$ for simplicity. We assume signals of length $N=4096$ containing pulses with a period length $T$ randomly chosen from a uniform distribution that ranges from 475 to 525, and random delay $\tau_0$ with constraint $0\leq \tau_0<T$. The resampling factor is $P_\text{R}=10$. The short pulses are chosen to be Gaussian pulses defined as $s_\text{Gauss}(t) =  \exp^{-\frac{t^2}{2\sigma^2}}$ with $\sigma = \frac{T_\text{p}}{6}$, and with $T_p$ as the desired pulse width (in our case $T_p=20$). These pulses were limited in length and are set to zero for $|t| > 3 \sigma$.
The \ac{ml} estimator in \eqref{equ:FD_Est_psi} that utilizes the multiharmonic signal model in \eqref{equ:MultHarmModel} requires the number of harmonic components $K_\text{h}$, which is strictly $K_\text{h} = \lfloor P/2 \rfloor$, whereas the power tends to decrease for higher harmonic components way below the noise floor. Thus, it is not guaranteed that the estimator performance increases with increasing model order. To find an appropriate model order for all \ac{snr} values, a model order selection has been implemented \cite[p.~40]{christensen2009multi}.

\begin{figure}[t]
	\centering
	\includegraphics[width=\columnwidth]{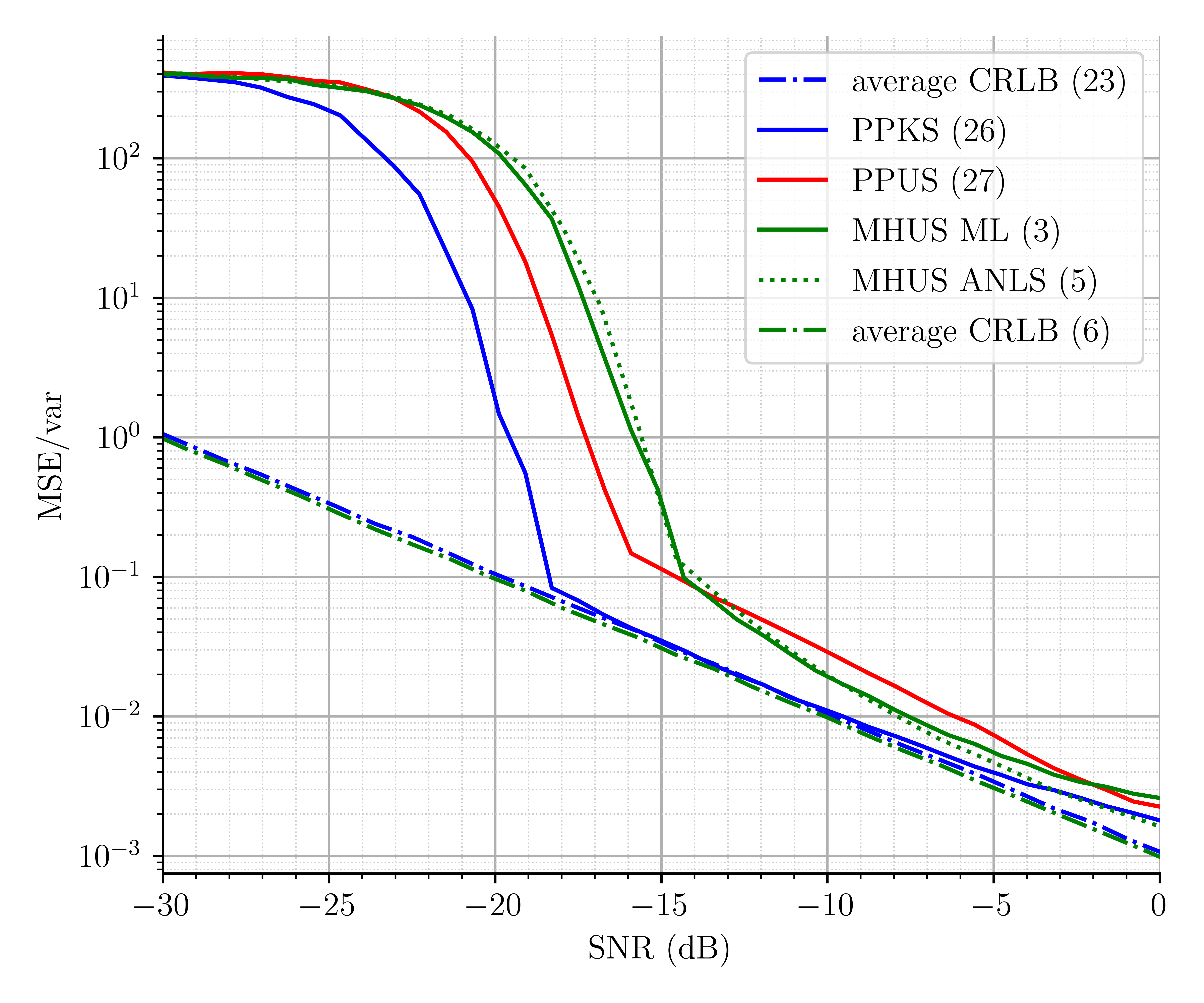}
	\caption{
    Simulated \acp{mse} of the \ac{ml} estimator for the period length $T$ with known (\ac{ppks}) and unknown (\ac{ppus}) information about the pulse shape as well as the \ac{anls} and \ac{ml} estimators based on the multiharmonic signal model (\ac{mhus}) with model order selection. The dash-dotted lines show the averaged \acp{crlb} in \eqref{equ:CRLB_MSB} and frequency-domain \eqref{equ:FD_CRLB_T} averaged over all realizations of $T$ and $\tau_0$.
    }
	\label{fig:MSE_curve}
\end{figure}

Fig.~\ref{fig:MSE_curve} shows the resulting \ac{mse} curves and \acp{crlb}. It can be seen that the presented estimators incorporating the knowledge of the short pulse duration offer in a better performance around the threshold. The threshold \ac{snr} for the best performing \ac{ppks} estimator is approximately $-18\,\text{dB}$, and it approximately attains the the \ac{crlb} above that. The \ac{ppus} and \ac{mhus} estimators don't have knowledge about the pulse shape, resulting in a deviation from the \ac{crlb}. Their overall performances are similar, while the \ac{ppus} estimator offers a better threshold \ac{snr}.

\section{Conclusion} \label{sec:Conclusions}
In this paper two signal models for short periodic pulses with and without known pulse shape were considered, and the corresponding \acp{fim} and closed-form approximations of them were derived. It turned out that the approximate \ac{fim} is singular in the case of an unknown pulse shape.
The \ac{ml} estimators for the signal period  were presented, which feature a better threshold behavior and better \ac{mse} performance for low \acp{snr} (and moderate in the case of the \ac{ppks} estimator) compared to estimator based on the multiharmonic signal model.

\section*{Acknowledgment}
This work was supported by Silicon Austria Labs (SAL), owned by the Republic of Austria, the Styrian Business Promotion Agency (SFG), the federal state of Carinthia, the Upper Austrian Research (UAR), and the Austrian Association for the Electric and Electronics Industry (FEEI).

\bibliographystyle{IEEEtran}
\bibliography{References}

\begin{thebibliography}{10}
\providecommand{\url}[1]{#1}
\csname url@samestyle\endcsname
\providecommand{\newblock}{\relax}
\providecommand{\bibinfo}[2]{#2}
\providecommand{\BIBentrySTDinterwordspacing}{\spaceskip=0pt\relax}
\providecommand{\BIBentryALTinterwordstretchfactor}{4}
\providecommand{\BIBentryALTinterwordspacing}{\spaceskip=\fontdimen2\font plus
\BIBentryALTinterwordstretchfactor\fontdimen3\font minus \fontdimen4\font\relax}
\providecommand{\BIBforeignlanguage}[2]{{%
\expandafter\ifx\csname l@#1\endcsname\relax
\typeout{** WARNING: IEEEtran.bst: No hyphenation pattern has been}%
\typeout{** loaded for the language `#1'. Using the pattern for}%
\typeout{** the default language instead.}%
\else
\language=\csname l@#1\endcsname
\fi
#2}}
\providecommand{\BIBdecl}{\relax}
\BIBdecl

\bibitem{hess2012pitch}
W.~Hess, \emph{Pitch Determination of Speech Signals: Algorithms and Devices}, ser. Springer Series in Information Sciences.\hskip 1em plus 0.5em minus 0.4em\relax Springer Berlin Heidelberg, 2012.

\bibitem{6850156}
M.~Pourhomayoun and M.~L. Fowler, ``Cramer-rao lower bound for frequency estimation for coherent pulse train with unknown pulse,'' \emph{IEEE Transactions on Aerospace and Electronic Systems}, vol.~50, no.~2, pp. 1304--1312, 2014.

\bibitem{Wise1976}
J.~Wise, J.~Caprio, and T.~Parks, ``Maximum likelihood pitch estimation,'' \emph{IEEE Transactions on Acoustics, Speech, and Signal Processing}, vol.~24, no.~5, pp. 418--423, 1976.

\bibitem{kay2017fundamentals}
S.~Kay, \emph{Fundamentals of Statistical Signal Processing, Volume 3}, ser. Fundamentals of statistical signal processing.\hskip 1em plus 0.5em minus 0.4em\relax Prentice Hall, 2017.

\bibitem{bang2023accurate}
J.-H. Bang, D.-H. Park, W.~Lee, D.~Kim, and H.-N. Kim, ``Accurate estimation of lpi radar pulse train parameters via change point detection,'' \emph{IEEE Access}, vol.~11, pp. 12\,796--12\,807, 2023.

\bibitem{kumar2014deinterleaving}
N.~U. Kumar, V.~Dhananjayulu, and V.~A. Kumar, ``Deinterleaving of radar signals and its parameter estimation in ew environment,'' \emph{International Journal of Emerging Technology and Advanced Engineering}, vol.~4, no.~9, pp. 490--494, 2014.

\bibitem{albaker2009signal}
B.~Albaker and N.~Rahim, ``Signal acquisition and parameter estimation of radio frequency pulse radar using novel method,'' \emph{IETE Journal of Research}, vol.~55, no.~3, pp. 128--134, 2009.

\bibitem{nishiguchi2000improved}
K.~Nishiguchi and M.~Kobayashi, ``Improved algorithm for estimating pulse repetition intervals,'' \emph{IEEE Transactions on Aerospace and Electronic Systems}, vol.~36, no.~2, pp. 407--421, 2000.

\bibitem{nishiguchi1983new}
K.~Nishiguchi, ``A new method for estimation of pulse repetition intervals,'' in \emph{National convention record of iece of japan}, 1983.

\bibitem{deng2012sfm}
B.~Deng, Y.-L. Qin, H.-Q. Wang, X.~Li, and L.~Nie, ``Sfm signal detection and parameter estimation based on pulse-repetition-interval transform,'' in \emph{European Signal Processing Conference (EUSIPCO)}, 2012, pp. 1855--1859.

\bibitem{mardia1989new}
H.~Mardia, ``New techniques for the deinterleaving of repetitive sequences,'' in \emph{IEE Proceedings F-Radar and Signal Processing}, vol. 136, no.~4.\hskip 1em plus 0.5em minus 0.4em\relax IET, 1989, pp. 149--154.

\bibitem{milojevic1992improved}
D.~Milojevi{\'c} and B.~M. Popovi{\'c}, ``Improved algorithm for the deinterleaving of radar pulses,'' in \emph{IEE Proceedings F (Radar and Signal Processing)}, vol. 139, no.~1.\hskip 1em plus 0.5em minus 0.4em\relax IET, 1992, pp. 98--104.

\bibitem{clarkson2008approximate}
I.~V.~L. Clarkson, ``Approximate maximum-likelihood period estimation from sparse, noisy timing data,'' \emph{IEEE transactions on signal processing}, vol.~56, no.~5, pp. 1779--1787, 2008.

\bibitem{logothetis1998interval}
A.~Logothetis and V.~Krishnamurthy, ``An interval-amplitude algorithm for deinterleaving stochastic pulse train sources,'' \emph{IEEE transactions on signal processing}, vol.~46, no.~5, pp. 1344--1350, 1998.

\bibitem{Williamson1994}
R.~C. Williamson, B.~James, B.~D. Anderson, and P.~J. Kootsookos, ``Threshold effects in multiharmonic maximum likelihood frequency estimation,'' \emph{Signal Processing}, vol.~37, 1994.

\bibitem{NIELSEN2017188}
\BIBentryALTinterwordspacing
J.~K. Nielsen, T.~L. Jensen, J.~R. Jensen, M.~G. Christensen, and S.~H. Jensen, ``Fast fundamental frequency estimation: Making a statistically efficient estimator computationally efficient,'' \emph{Signal Processing}, vol. 135, pp. 188--197, 2017. [Online]. Available: \url{https://www.sciencedirect.com/science/article/pii/S0165168417300117}
\BIBentrySTDinterwordspacing

\bibitem{nehorai1986adaptive}
A.~Nehorai and B.~Porat, ``{Adaptive comb filtering for harmonic signal enhancement},'' \emph{IEEE Transactions on Acoustics, Speech, and Signal Processing}, vol.~34, no.~5, pp. 1124--1138, 1986.

\bibitem{10476910}
J.~Lindenberger, S.~Schuster, O.~Lang, A.~Haberl, C.~Staudinger, T.~Roland, and M.~Huemer, ``Bias, variance, and threshold level of the least squares pitch estimator with windowed data,'' in \emph{2023 57th Asilomar Conference on Signals, Systems, and Computers}, 2023, pp. 854--859.

\bibitem{CooleyTukey}
J.~Cooley and J.~Tukey, ``An algorithm for the machine calculation of complex fourier series,'' \emph{Mathematics of Computation}, vol.~19, no.~90, pp. 297--301, 1965.

\bibitem{Kay-Est.Theory}
S.~M. Kay, \emph{{Fundamentals of Statistical Signal Processing: Estimation Theory}}.\hskip 1em plus 0.5em minus 0.4em\relax Prentice Hall, 1993, vol.~1.

\bibitem{MR15748}
C.~Radhakrishna~Rao, ``Information and the accuracy attainable in the estimation of statistical parameters,'' \emph{Bull. Calcutta Math. Soc.}, vol.~37, pp. 81--91, 1945.

\bibitem{cramér1946mathematical}
H.~Cram{\'e}r, \emph{Mathematical Methods of Statistics}, ser. Goldstine Printed Materials.\hskip 1em plus 0.5em minus 0.4em\relax Princeton University Press, 1946.

\bibitem{Hoerl01021970}
A.~E. Hoerl and R.~W. Kennard, ``Ridge regression: Biased estimation for nonorthogonal problems,'' \emph{Technometrics}, vol.~12, no.~1, pp. 55--67, 1970.

\bibitem{christensen2009multi}
M.~Christensen and A.~Jakobsson, \emph{{Multi-Pitch Estimation}}.\hskip 1em plus 0.5em minus 0.4em\relax Morgan \& Claypool Publishers, 2009.

\end{thebibliography}
\end{document}